\documentclass{optica-article}
\usepackage{booktabs} 
\usepackage{float}
\usepackage{svg}

\journal{opticajournal} 

\articletype{Research Article}

\begin{document}

\title{Quantum Random Number Generation Based on Phase Reconstruction}

\author{Jialiang Li,\authormark{1,2} Zitao Huang,\authormark{1,2} Chunlin Yu,\authormark{3} Jiajie Wu,\authormark{3} Tongge Zhao,\authormark{3} Xiangwei Zhu,\authormark{1,2} and Shihai Sun\authormark{1,2,*} }

\address{\authormark{1}Shenzhen Campus of Sun Yat-sen University, No. 66, Gongchang Road, Guangming District, Shenzhen, Guangdong 518107, P.R. China\\
\authormark{2}School of Electronics and Communication Engineering, Sun Yat-sen University, Shenzhen, Guangdong 518107, P.R. China\\
\authormark{3}China Greatwall Research Institute, China Greatwall Technology Group CO., LTD., Shenzhen, Guangdong 518107, P.R. China
}

\email{\authormark{*}sunshh8@mail.sysu.edu.cn} 


\begin{abstract*} 
Quantum random number generator (QRNG) utilizes the intrinsic randomness of quantum systems to generate completely unpredictable and genuine random numbers, finding wide applications across many fields. QRNGs relying on the phase noise of a laser have attracted considerable attention due to their straightforward system architecture and high random number generation rates. However, traditional phase noise QRNGs suffer from a 50\% loss of quantum entropy during the randomness extraction process. In this paper, we propose a phase-reconstruction quantum random number generation scheme, in which the phase noise of a laser is reconstructed by simultaneously measuring the orthogonal quadratures of the light field using balanced detectors. This enables direct discretization of uniform phase noise, and the min-entropy can achieve a value of 1. Furthermore, our approach exhibits inherent robustness against the classical phase fluctuations of the unbalanced interferometer, eliminating the need for active compensation. Finally, we conducted experimental validation using commercial optical hybrid and balanced detectors, achieving a random number generation rate of 1.96 Gbps at a sampling rate of 200 MSa/s.

\end{abstract*}

\section{Introduction}
Random numbers play a crucial role in various applications, particularly in cryptography \cite{c1,c2,c3,c4,c5}. Currently, two primary methods are employed for generating random numbers. The first method utilizes pseudo-random number generators based on computer algorithms, such as linear congruential generators and Mersenne Twister generators \cite{c6,c7}, to generate longer random number sequences from shorter seed numbers through deterministic algorithms. Pseudo-random number generators are simple to implement and can generate random number sequences of any length with minimal computational cost. However, these sequences are completely predictable for a given seed number and expansion algorithm, making them inappropriate for secure applications like cryptography. The second method involves random number generators based on classical physical noise, which generate random numbers by measuring real physical quantities of complex systems, such as random number generators based on chaotic events and mouse movement \cite{c8,c9}. In comparison to pseudo-random numbers, random numbers derived from classical physical noise offer improved unpredictability and security. However, classical physics is founded on deterministic principles in theory, achieving randomness through incomplete parameter descriptions. In principle, an adept eavesdropper may predict random numbers through device defects and other side-channel attacks.

Fortunately, quantum random number generators (QRNGs) have been proposed based on quantum superposition and probabilistic measurements, which can achieve genuinely unpredictable random numbers. QRNGs have been implemented with the path or polarization of photon \cite{c10,c11}, photon arrival time \cite{c12,c13,c14}, photon numbers \cite{c15,c16}, etc., which are simple in principle but have low generation rates. Researchers have proposed high-speed structures for random number generation utilizing macroscopic physical properties, including vacuum fluctuations \cite{c17,c18,c52}, phase noise \cite{c19,c20,c21,c22,c23,c49,c51}, and amplified spontaneous emission noise \cite{c24,c25}, which can achieve quantum random number generation speeds of up to 100 Gbps. Furthermore, quantum random number chips based on integrated photonic circuits have been extensively researched for addressing the bottlenecks such as volume, power consumption, and stability \cite{c30,c47,c48}.

Ideal QRNGs can generate genuine random numbers with full entropy in theory. However, practical QRNGs face challenges due to two constraints associated with their randomness.  First, they are naturally affected by classical noise in measurement systems, such as electrical noise from quantum measuring devices. Although careful entropy assessment and subsequent data post-processing can eliminate this classical noise, ensuring the randomness of generated data originates solely from quantum noise. Challenges persist in developing real-time, cost-effective post-processing hardware, particularly for high-speed QRNGs. Second, QRNGs can be influenced by parameters hidden from eavesdroppers, allowing them to predict the sequence. Researchers proposed device-independent QRNGs \cite{c26,c27,c28} to mitigate this issue, where security is ensured by violating Bell inequalities in quantum systems. The disadvantages of low random number generation rate and increased system complexity limit the practical application of this method. Besides, many alternative approaches have been developed, such as source-independent QRNGs \cite{c29,c30} and measurement-device-independent QRNGs \cite{c31,c32}, to strike a balance between security and random number generation rate.

Due to simple implementation and high rate of random number generation, QRNG based on phase noise has garnered significant attention and research interest \cite{c19,c33,c34}. The phase in the output field of the laser is influenced by the randomness of spontaneous emission photons and conforms to a Gaussian distribution \cite{c19,c35}, with a variance represented by
\begin{equation}
\sigma^2 = \langle \Delta \xi(t) \rangle^2 = \frac{2 T_{\rm delay}}{\tau_c}
\end{equation}
where $\tau_c$ represents the coherence time of the laser, inversely proportional to its linewidth $\tau_c \approx 1/(\pi \Delta f)$. The time delay is $T_{\rm delay}=n \Delta L/c$, where $n$ is the refractive index of the optical fiber, $c$ is the speed of light in a vacuum, and $\Delta L$ is the length of the delay line. 

The phase $\Delta \phi_0(t)$ is a variable that maps $\Delta \xi(t)$ to the range $[- \pi, \pi)$, following a folded Gaussian distribution \cite{c36,c37}. The distribution of $\Delta \phi_0(t)$ can be expressed as
\begin{equation}
f_{\Delta \phi_0} (\Delta \xi) =\frac{1}{\sigma \sqrt{2\pi}} \sum^{+\infty}_{k=-\infty} \exp[-\frac{(\Delta \xi-2k\pi)^2}{2\sigma^2}]
\end{equation}

Random number extraction can be achieved through an unbalanced interferometer that measures the phase of the laser. Some conditions need to be satisfied to attain the phase noise correctly: a) $T_R<\tau_c$, ensuring the detector can effectively capture the phase noise; b) $T_d \gg \tau_c$, guaranteeing a uniform distribution of the phase noise, where $T_R$ is the response time, $\tau_c$  is the coherence time of the laser, $T_d$ is the delay introduced by the arm length difference of the interferometer, and $T_S$ is the sampling period.

Although QRNG based on the unbalanced interferometer and direct measurement using photodiode (PD) has the advantage of simple structure, it still faces two shortcomings in practical systems. One primary concern arises from environmental factors, such as vibration and temperature, where unbalanced interferometer systems experience phase drift, introducing additional classical phase fluctuations. Many studies have been presented to alleviate this problem. Qi et al. proposed to stabilize the unbalanced interferometer with active controllers \cite{c19}. Xu et al. implemented an internally temperature-controlled compact planar lightwave circuit (PLC) unbalanced interferometer, achieving phase stabilization through temperature control \cite{c23}. Nie et al. by using a polarization-insensitive Michelson interferometer and active PID algorithms, improved the stability of the interferometer \cite{c38}. In addition, one method utilizes two independent lasers for interference that occurs within a multimode interference (MMI) device to avoid the influence of an unbalanced interferometer, while ensuring that the phase noise follows a uniform distribution \cite{c39,c40,c41,c51}. However, two lasers with the same center frequency and spectral characteristics, as well as additional temperature control modules, are required to ensure the stability of the laser wavelength.

The second concern pertains to the quantum entropy loss in the post-processing of sampled data. The quantum min-entropy quantifies the maximum randomness extracted from a single sample, which is very low in traditional phase noise QRNGs \cite{c20,c39,c42}. The quantum min-entropy is about 0.5 for the phase noise measured with an unbalanced interferometer and PD, which implies that a lot of the original bits will be lost, severely limiting the quantum random number generation rate.

In this paper, a QRNG based on phase reconstruction is proposed, which achieves the recovery of the phase information by simultaneously measuring the orthogonal quadratures of the optical field using balanced detectors, enabling direct discretization of phase noise. In contrast to traditional phase measurement methods, our proposed approach offers several advantages: quantum entropy close to 1 is achieved by employing post-processing techniques on the sampled data, leading to a higher rate of random number generation under identical conditions. Our approach is insensitive to the phase fluctuations of an unbalanced interferometer, exhibiting robustness against environmental factors and eliminating the need for intricate phase stabilization measures. Furthermore, a comprehensive model is established to analyze the imperfection of practical devices and the influence of classical noise. Finally, we validated our approach through experiments using commercially available components. The proposed achieves a high output rate of 1.96 Gbps at a sampling rate of 200 MSa/s, successfully passing the rigorous NIST random tests.

\section{Theory}
\subsection{Extraction of phase noise}

\begin{figure}[hbp]
\centering
\includegraphics[width=13cm]{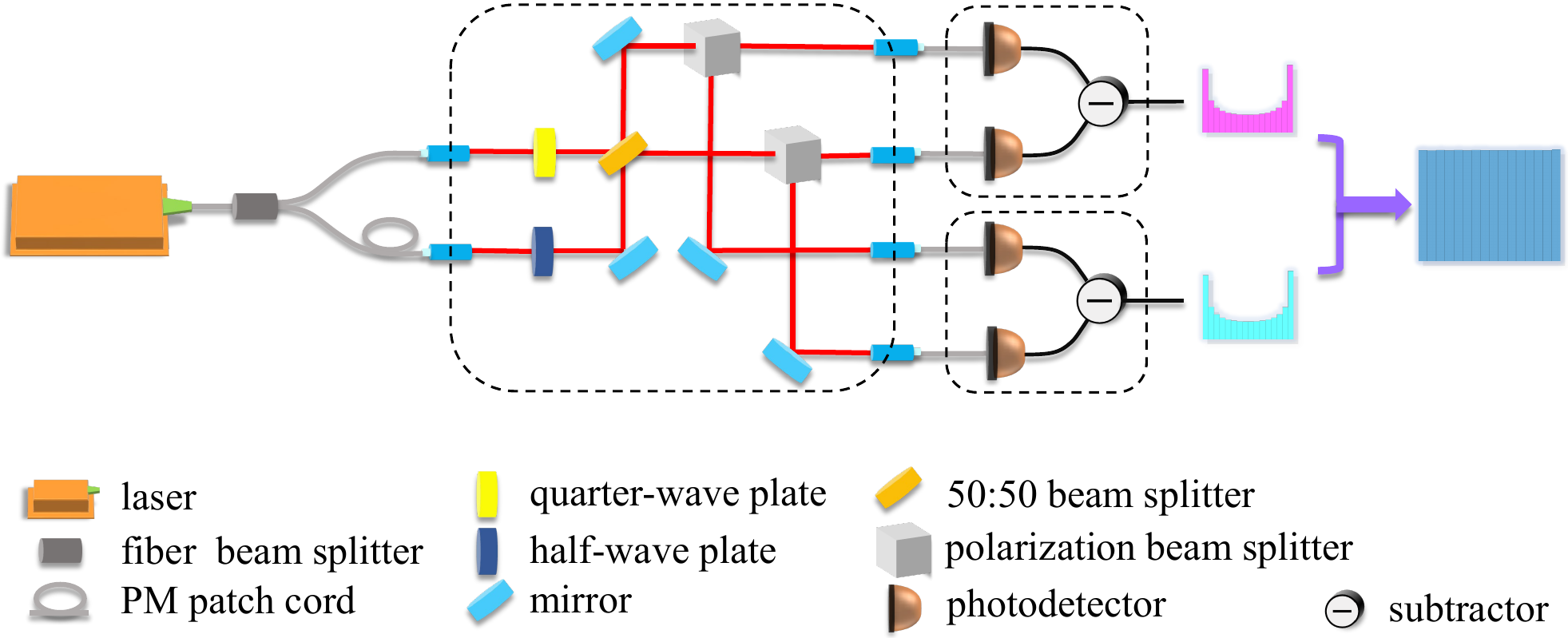}
\caption{Principle diagram of phase-reconstruction QRNG.}
\end{figure}

The phase reconstruction QRNG scheme is depicted in Fig. 1. A continuous wave emitted by the laser is divided into two beams, one beam serves as the local oscillator light (LO), while the other functions as the signal light after a delay line. The local oscillator light passes through a quarter-wave plate oriented at 45 degrees relative to the LO polarization direction, converting into circularly polarized light and generating two polarization components with a relative 90 degrees phase delay orthogonal to each other. The signal light traverses through a half-wave plate set at 22.5 degrees relative to the signal polarization direction, producing two polarization components orthogonal to each other. In the LO and signal light components, those with matching polarization orientations interfere at the beam splitter (BS), and the resulting interference signals under different polarizations are split via polarization beam splitters (PBSs). Then, the two components of the optical field, $I$ and $Q$, are measured by two balanced homodyne detectors (BHDs) to recover the phase information \cite{c43}.

The field of a laser output can be described as \cite{c19,c44,c45}
\begin{equation}
E_0(t)=\sqrt{P_0} \exp[j(\omega_0 t+\phi_0(t))],
\end{equation}
where $P_0$ represents the optical power output, $\omega_0$ represents the laser angular frequency, and $\phi_0(t)$ represents the instantaneous phase of the laser.

The output interference signal $V_I$  and $V_Q$ of two BHDs are expressed as
\begin{equation}
\begin{cases}
V_I=Z_1 R_1\sqrt{P_{\rm S} P_{\rm LO}} \cos (\omega_0 T_{\mathrm{delay}}+\Delta\phi_0(t)) \equiv I_0 \cos(\Delta\phi(t)),\\
V_Q=Z_2 R_2\sqrt{P_{\rm S} P_{\rm LO}} \sin (\omega_0 T_{\mathrm{delay}}+\Delta\phi_0(t)) \equiv Q_0 \sin(\Delta\phi(t)).
\end{cases}
\end{equation}
where $Z_1$, $Z_2$, $R_1$ and $R_2$ represent the transimpedance gain and responsivity of two BHDs. $P_{\rm S}$ and $P_{\rm LO}$ are the power of signal light and local oscillator light. $\omega_0 T_{\mathrm{delay}}$ is the inherent phase difference caused by the delay $T_{\mathrm{delay}}$, and $\Delta\phi_0(t)$ is the instantaneous phase difference between the signal light and the local oscillator light. According to the complex expression of the optical field $V_Z=V_I+iV_Q=|V_Z| \exp(i\Delta \phi)$, where $|V_Z|=\sqrt{V_I^2+V_Q^2}$, the phase information of the optical field can be obtained as
\begin{equation}
\Delta\phi=\arctan{\frac{V_Q}{V_I}}.
\end{equation} 

Figure 2 shows the distribution of $\Delta\phi(t)$ under various phase noise variances. As the variance of phase noise steadily increases from 0, the statistical distribution of the reconstructed phase transitions successively from a Gaussian distribution to a truncated Gaussian distribution and ultimately to a uniform distribution. Two distinct critical variances, 0.6 and 10, play pivotal roles as thresholds delineating pivotal stages in the evolution of the statistical characteristics of the reconstructed phase. Specifically, the critical variance of 0.6 signifies the transition from a Gaussian to a truncated Gaussian distribution, while the critical variance of 10 demarcates the shift from a truncated Gaussian distribution to a uniform distribution. 

The following conclusions can be drawn: (a) When the variance $\sigma^2<0.6$, $\Delta\phi(t)$ follows the same Gaussian distribution as $\Delta\xi(t)$. Interference signal $V_I$ is mainly distributed around the maximum value of 1, and the distribution of $V_Q$ is concentrated on the value of 0, and the distribution is symmetric. (b) When the variance $0.6\le\sigma^2<10$, $\Delta\phi(t)$ exhibits a truncated Gaussian distribution. The distributions of $V_I$ and $V_Q$ both take on a U-shaped form, with different probabilities around the maximum value of 1 and the minimum value of -1, resulting in an asymmetric distribution. (c) When the variance $\sigma^2\geq10$, $\Delta\phi(t)$ approaches a uniform distribution. Both $V_I$ and $V_Q$ exhibit arcsine distributions with the same parameters, symmetrically centered around the value of 0. From the above analysis, it can be seen that there exists a critical variance $\sigma_0^2=10$.

\subsection{The imperfections of practical devices}
In ideal situations (see Fig. 4(a)), the complex information of the interference wave can be accurately reconstructed. However, in practical experiments, device imperfections and classical noise inevitably influence the randomness of the generated bits. Hence, we give a detailed analysis to take the main imperfections into account and evaluate the performance of our method.

\textbf{Splitting ratio of BS-} By considering the transmittance $T$ of BS and the delay line loss coefficient $K$ as shown in Fig. 1, $P_{\rm S}=KTP_0$ and $P_{\rm LO}=(1-T)P_0$. The amplitudes of the output signals from the BHDs are rewritten as
\begin{equation}
\begin{cases}
I_0=Z R P_0 \sqrt{KT(1-T)},\\
V_0=Z R P_0 \sqrt{KT(1-T)}.
\end{cases}
\end{equation}

Eq. (6) reveals that the output signals from BHDs reach maximum amplitude when BS has a 50:50 splitting ratio ($T=1/2$). Yet, the transmittance $T$ is not equal to $1/2$, thus the outputs $V_I$ and $V_Q$ from the BHD are proportionally reduced, leading to a reduction in the amplitude of the complex information $V_Z$ while the phase remains unchanged. However, a significant difference in splitting ratio leads to a rapid decline in the amplitudes of the output orthogonal quadratures, thereby reducing interference visibility. This situation is unfavorable for observing and acquiring signals.

\begin{figure}[htbp]
\centering
\includegraphics[width=13cm]{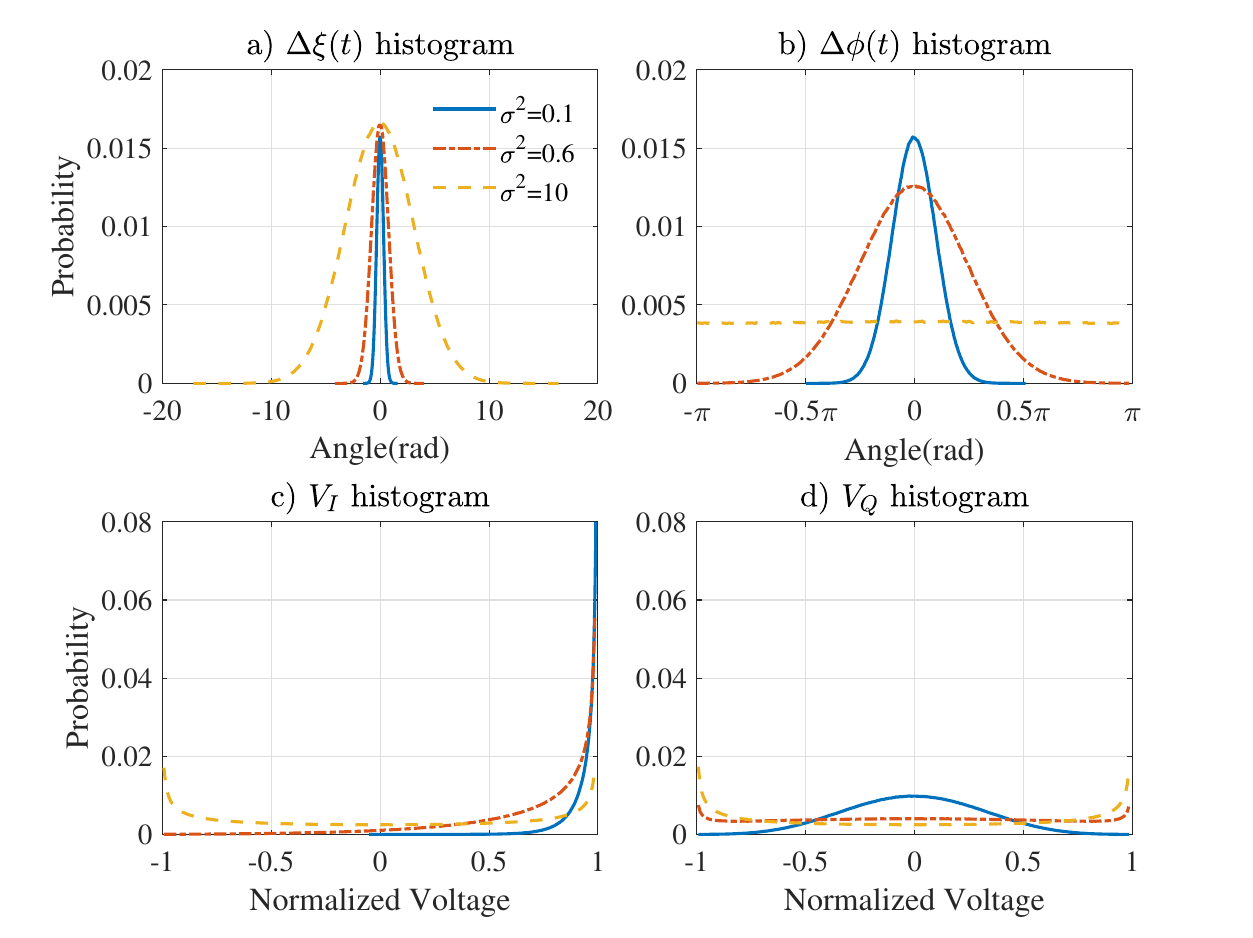}
\caption{The impact of noise variance on the output distribution. The changes in the output distribution under three different noise variances 0.1, 0.6, 10 are analyzed. (a) Distribution of the real phase noise of the laser. (b) Distribution of the phase mapped to $[-\pi,\pi)$. (c) Distribution of the $I$ component. (d) Distribution of the $Q$ component.}
\end{figure}

\textbf{Unmatched BHDs-} In the discussion in the previous section, a basic assumption is that $Z_1=Z_2$ and $R_1=R_2$, therefore, the reconstructed phase $\varphi$ is the true phase of the complex information (see Eq. (5)). But the scenario where $I_0\neq Q_0$ is a common occurrence in practical settings due to inconsistency of the gain or responsivity, in which both the amplitude and phase of the complex information $V_Z$ will change and an additional phase will be introduced. The greater the difference between $I_0$ and $Q_0$, the larger the additional phase (see Fig. 3).

Nevertheless, the difference between $I_0$ and $Q_0$ remains consistent across both detection channels. Therefore, by measuring the disparity between $I_0$ and $Q_0$, the system can realize the extraction of the true phase from the measured values. Assuming measured values is $V_I^\prime=I_0^\prime \cos\varphi^\prime$ and $V_Q^\prime=Q_0^\prime \sin\varphi^\prime$, the additional phase can be expressed as
\begin{equation}
\Delta\varphi=\varphi^\prime-\varphi=\arctan{\frac{I_0^\prime Q_0-I_0 Q_0^\prime}{Q_0 Q_0^\prime+I_0^\prime I_0}}.
\end{equation}

\begin{figure}[htbp]
\centering
\includegraphics[width=10cm]{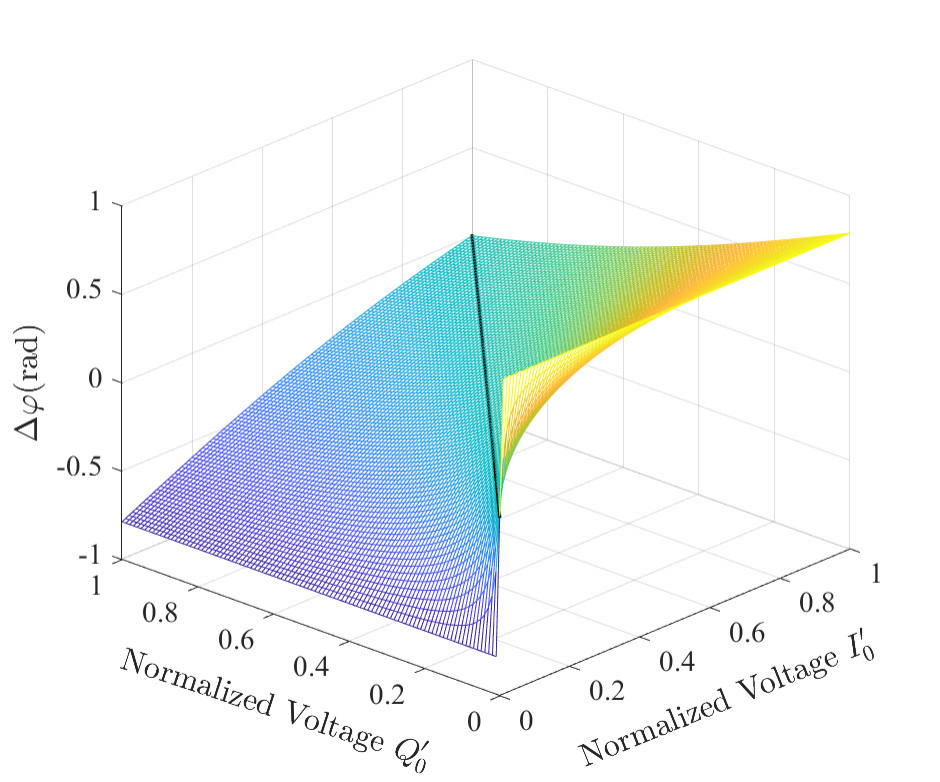}
\caption{Variation of the additional phase with the amplitude of the orthogonal quadratures. $I_0^\prime$ and $Q_0^\prime$ are the practical normalized amplitudes of the two BHDs. The black line represents the case when the additional phase is zero. It can be observed that when $I_0^\prime>Q_0^\prime$, the additional phase is positive; when $I_0^\prime<Q_0^\prime$, the additional phase is negative.}
\end{figure}

\textbf{Intensity fluctuations of light-} In practical systems, the laser’s actual output power fluctuates over time, denoted as the mean $\overline{P_0}$ combined with intensity fluctuations $\epsilon(t)$. Therefore, both the local oscillator light and the signal light should be expressed as
\begin{equation}
\begin{cases}
P_{\rm S}(t)=\overline{P_{\rm S}} + \epsilon_{\rm S}(t),\\
P_{\rm LO}(t)=\overline{P_{\rm LO}} + \epsilon_{\rm LO}(t).
\end{cases}
\end{equation}
where we assume that the intensity fluctuations $\epsilon_{\rm S,\rm LO}(t)$ follow a zero-mean Gaussian white noise with variances $\sigma_{\epsilon_{\rm S,\rm LO}}^2$, and $\epsilon_{\rm S}(t)$, $\epsilon_{\rm LO}(t)$ are mutually independent. 

Figure 4(b) shows the probability distribution when only the intensity fluctuations are considered. It shows that the intensity fluctuations result in a sharp decrease in the right-side peak of the distribution, with a slight decrease in the left-side peak. Generally, the effect of intensity fluctuations primarily manifests as the smoothing of the positive voltage peak, which disrupts the original symmetry of the output distribution.

\textbf{Electrical noise of the detectors-} Classical electrical noise within the detection apparatus primarily comes from the photodetector and oscilloscope. It is generally assumed that both of these noise sources are independent Gaussian white noises. Thus, the cumulative electrical noise contributes Gaussian white noise to the signal. This can be expressed as
\begin{equation}
V(t)=V_0 \cos(\Delta \phi)+w(t).
\end{equation}

Figure 4(c) shows the affection of the electrical noise, which has a smoothing effect on the voltage distribution. However, in contrast to intensity noise, electrical noise affects both positive and negative peak values in the same manner. Consequently, the distribution of output voltage remains symmetric. The larger the variance of the electrical noise, the more pronounced the smoothing effect, leading to a more rapid decrease in peak values in the output distribution.

\begin{figure}[htbp]
\centering
\includegraphics[width=14cm]{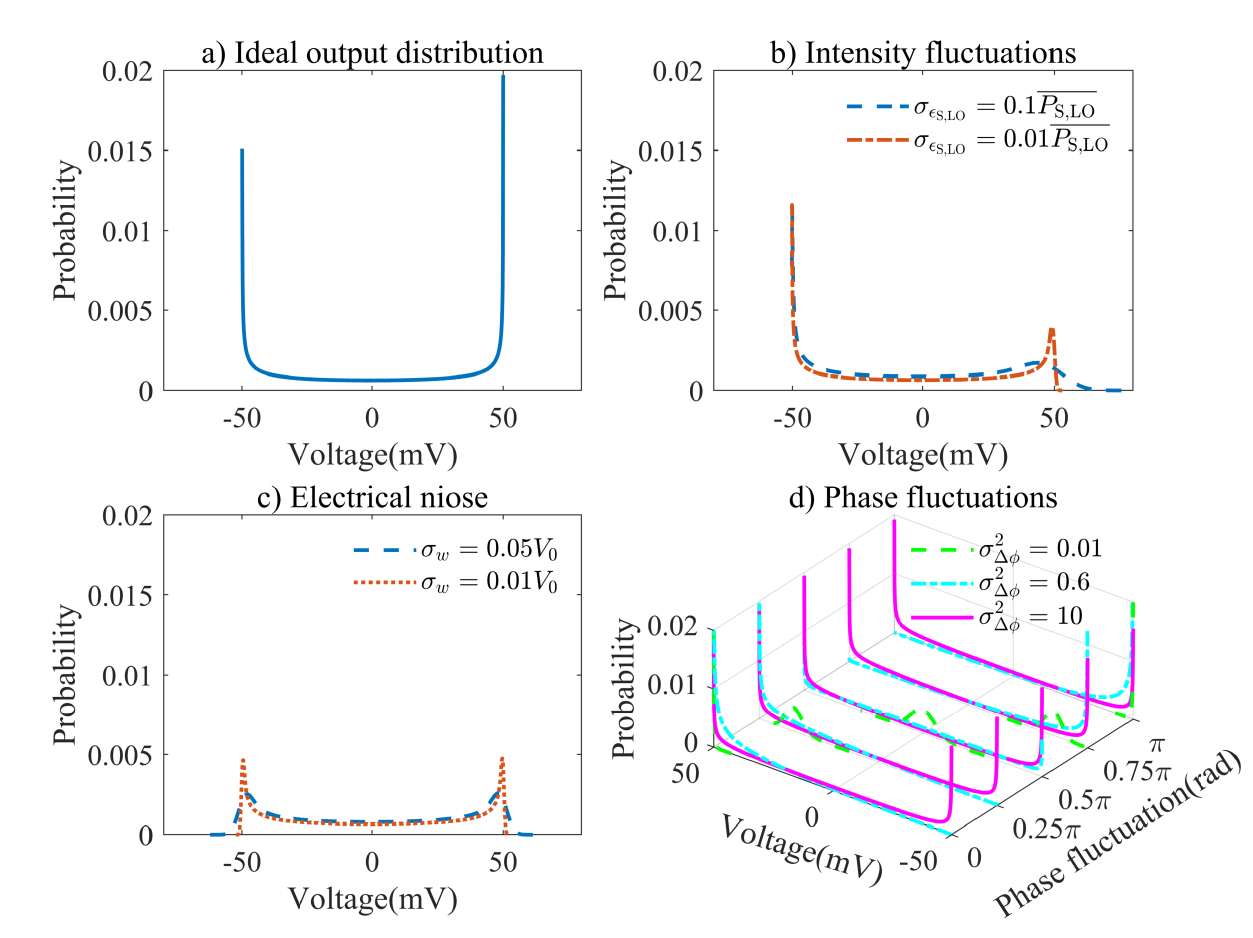}
\caption{Impact of classical noise on the distribution of output voltage. The distribution of output voltage under four scenarios is described (a) in the absence of classical noise, (b) with only intensity fluctuations, (c) with only electrical noise, and (d) with only classical phase fluctuations. }
\end{figure}

\textbf{Stability of unbalanced interferometer-} An unbalanced interferometer should be introduced to measure the phase noise of a laser, but the interferometer is inevitably affected by environments (such as temperature), which will introduce classical phase fluctuations to the interferometer \cite{c25}. Generally speaking, the variation of classical phase fluctuations is significantly slower than the sampling period. It can be regarded as a random variable because its value varies over different sampling periods but remains constant within one sampling period. The output voltage can be expressed as
\begin{equation}
V(t)=V_0 \cos(\Delta \phi + \phi_0).
\end{equation}

Figure 4(d) shows the distribution of output voltage with different phase fluctuations $\phi_0$, which are systematically chosen within the interval $[0,\pi]$. When the variance of $\Delta\phi$ is greater than the critical variance $\sigma_0^2$, classical phase fluctuations have almost no impact on the output voltage distribution.

\textbf{Total noise of system-} In practical scenarios, the system is subjected to the influence of multiple factors, rather than being solely impacted by a single factor. Hence, accounting for all the practical factors, the expression for the output voltage is
\begin{equation}
V(t)=Z R \sqrt{[\overline{P_{\rm S}}+\epsilon_{\rm S}(t)][\overline{P_{\rm LO}}+\epsilon_{\rm LO}(t)]} \cos(\Delta \phi+\phi_0)+w(t).
\end{equation}

\begin{figure}[htbp]
\centering
\includegraphics[width=10cm]{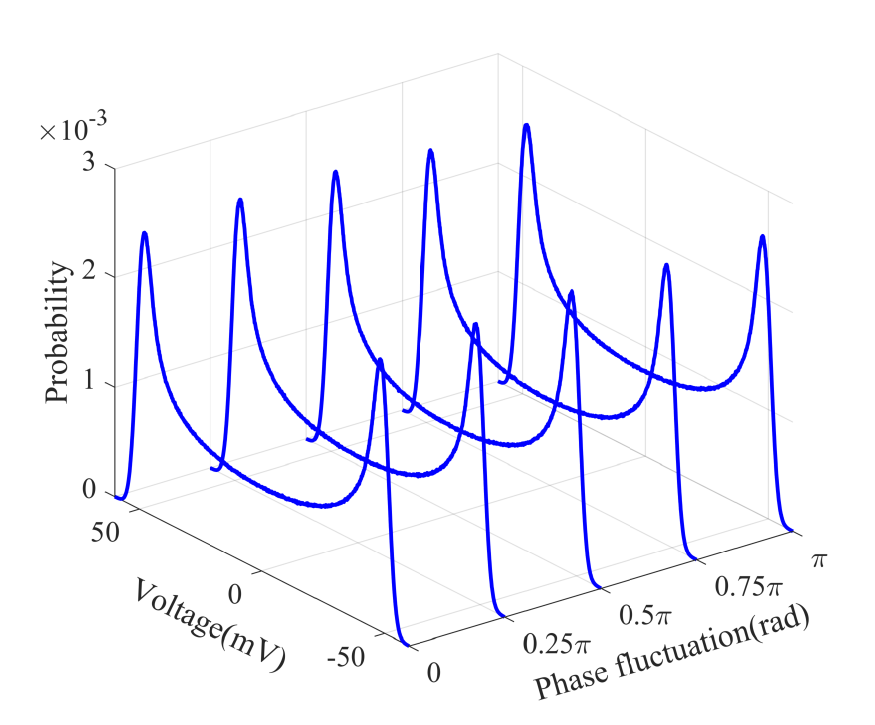}
\caption{Distribution of output voltage under the combined influence of all classical noise.}
\end{figure}

In instances where the variance of $\Delta\phi$ exceeds the critical value $\sigma_0^2$, with intensity fluctuations having a variance of $\sigma_{\epsilon_{\rm S, \rm LO}}=0.01\overline{P_{\rm S, \rm LO}}$, and electrical noise exhibiting a variance of $\sigma_w=0.1V_{\max}$, the outcomes are depicted in Fig. 5. Under these conditions, the output distribution remains unaffected by phase fluctuations. The amplitudes on both sides of the distribution exhibit near-identity, suggesting minimal influence from intensity fluctuations, with electrical noise assuming a dominant role. In the event of an increase in the variance of intensity fluctuations or electrical noise, both peaks of the distribution undergo a uniform smoothing effect, maintaining the symmetry of the distribution. In instances where the phase noise variance is less than the critical value $\sigma_0^2$, significant alterations occur in the voltage distribution owing to the presence of phase fluctuations. Under these circumstances, the stability of the output distribution is affected, requiring additional measures to maintain stability.

The presence of classical noise, particularly phase fluctuations in the interferometer, renders the output of QRNG unstable. Hence, our QRNG scheme exhibits robustness against classical phase fluctuations in the unbalanced interferometer and the system demonstrates sustained and stable output over extended periods.

\section{Experimental setup and results}
The experimental setup is illustrated in Fig. 6. A continuous-wave optical signal is sent by a distributed feedback (DFB) laser with a wavelength of 1550 nm. The optical signal is split into two beams by a BS, one beam serves as the local oscillator light, and the other serves as the signal light after a 6m polarization maintaining (PM) fiber. The local oscillator light and the signal light are directed into a 90-degree optical hybrid (Optoplex HB-C0AFAC057), thereby creating four orthogonal states within the complex field space. These output optical signals are input into two BHDs (Thorlabs PDB480C-AC) for photoelectric conversion. Subsequently, the two output signals ($I$ and $Q$) are acquired and quantified by using a high-speed oscilloscope (Keysight Infiniium DSOS104A) with a bandwidth of 1 GHz and a sampling rate of 20 GSa/s.

\begin{figure}[htbp]
\centering
\includegraphics[width=14cm]{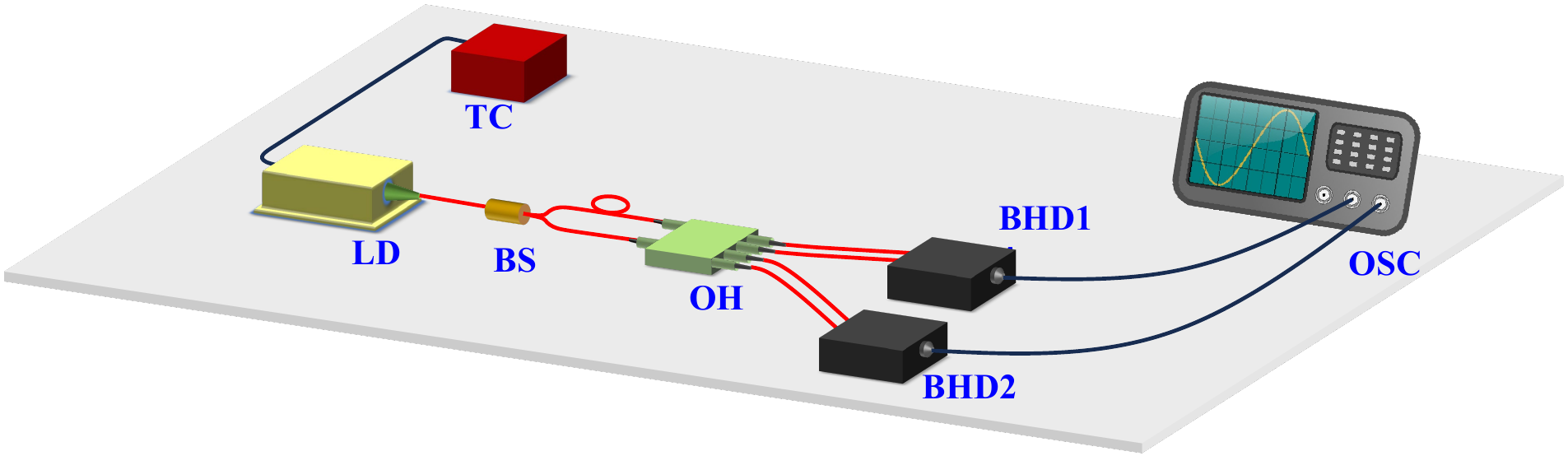}
\caption{Experimental setup of the phase-reconstruction QRNG. A stable continuous wave is emitted by a laser driven by a temperature controller (TC), divided into two paths by a BS and a delay line, and input into an optical hybrid (OH) to generate four orthogonal states. These states are then detected by two BHDs and signal acquisition and 10-bit quantization is performed using an oscilloscope (OSC).}
\end{figure}

The linewidth of the laser is about 50 MHz (coherence time of 6 ns). And the laser diode is driven by a compact laser diode controller (Thorlabs CLD1015) to stabilize its power and wavelength in real time. The controller settings are optimized to maintain the laser diode temperature at 25°C and the drive current at 14.5 mA. This configuration yields an output power of 0.140 mW with a standard deviation of power fluctuations at $7.855\times{10}^{-7}$ W. The practical output powers of BS are about 67.5 $\mu$W and 64.5 $\mu$W at two beams respectively. The balanced detectors have a bandwidth of 1.6 GHz, a response time of 625 ps, a responsivity of $R$=1 A/W, and a transimpedance gain of $Z=16\times{10}^3$ V/A. The output voltages amplitude of the BHDs are both approximately 600 mV.

According to the analysis in Section 2.1, the orthogonal quadratures $I$ and $Q$ from the detectors follow an arcsine distribution. Thus, the probability density function can be expressed as
\begin{equation}
f_X(x) = \begin{cases}
\frac{1}{\pi\sqrt{A^2-x^2}}, &-A<x<A\\
0, &else	
\end{cases}
\end{equation}
here $X=I$ or $Q$, $A$ is the maximum amplitude of $X$. The actual statistical distributions of $I$ and $Q$ are shown in Fig. 7(a) and Fig. 7(b). In this experiment, the power of the local oscillator $P_{\rm LO}$ is 0.068 mW with a standard deviation of $4.505\times{10}^{-7}$ W, while the power of the signal light $P_{\rm S}$ is 0.041 mW with a standard deviation of $2.823\times{10}^{-7}$ W. The standard deviations of the electronic noise for the $I$ and $Q$ channels are $7.666\times{10}^{-3}$ V and $7.356\times{10}^{-3}$ V\ respectively. Submitting all the parameters into Eq. (9), we could evaluate the performance of our QRNG system.

Figures 7(a) and (b) illustrate both the simulated and experimental results for both $I$ and $Q$. In the measured data, the amplitudes of the bias-removed $I$ and $Q$ are 595.5 mV and 567.5 mV respectively, which are much larger than the electrical noise of two BHDs with 41.1 mV and 40.3 mV. To reduce the affection of the unmatched gains between two BHDs, the output voltage of $I$ and $Q$ have been normalized. The results show that the experimental data are matched with that of the simulation, validating that our model can predict the output of the practical QRNG system.

\begin{figure}[htbp]
\centering
\includegraphics[width=13cm]{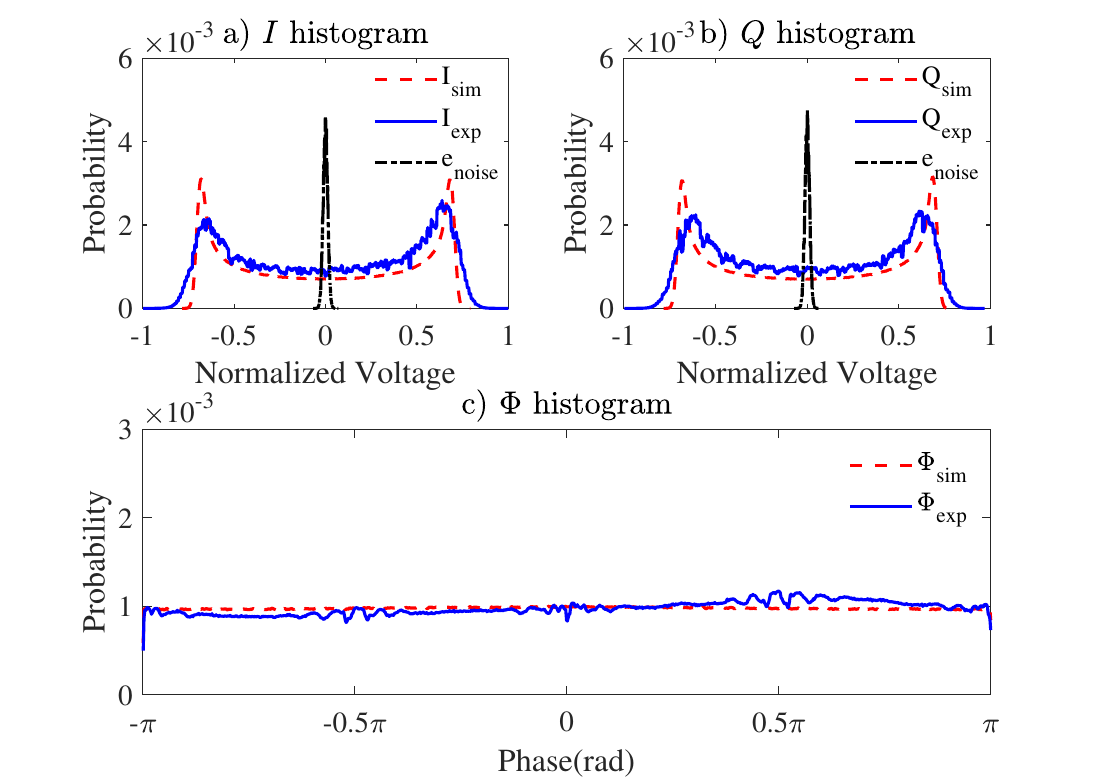}
\caption{Distribution of simulation and experimental data. The red dashed line in the figure represents simulation data, the black dash-dot line represents electrical noise, and the blue solid line represents experimental data. (a) Distribution of the $I$ component. (b) Distribution of the $Q$ component. (c) Distribution of the reconstructed phase.}
\end{figure}

With the measured $I$ and $Q$, we could reconstruct the phase of light, which can be written as
\begin{equation}
\Phi=\arg (I+iQ),
\end{equation}
here $\arg$ denotes the phase of a complex number. The phase $\Phi$ exhibits a uniform distribution within the range $[-\pi,\pi)$, with its probability density function expressed as
\begin{equation}
    f_{\Phi}(\phi)=\begin{cases}
        \frac{1}{2\pi}, &-\pi \leq \phi < \pi\\
        0, &else
    \end{cases}
\end{equation}

Following the acquisition of orthogonal quadratures, the phase undergoes quantization into $2^n$ bins, each with a width of $\Delta=\frac{\pi}{2^{n-1}}$. The probability in each bin is the same, with the probability of the first bin calculated as
\begin{equation}
   P_{\Phi}(\phi=\phi_1)=\int^{-\pi+\frac{\pi}{2^{n-1}}}_{-\pi} f_{\Phi}(\phi)d\phi=\frac{1}{2^n}
\end{equation}

To evaluate the randomness of the samples, the min-entropy ($H_{\min}$) is commonly employed \cite{c50}. For a random variable $X$, the quantum min-entropy is defined as
\begin{equation}
H_{\rm min}(X)=-\log_2[\max_{x_i\in X}P(X=x_i)].
\end{equation}
where $x_i$ signifies an element of $X$, and $P_X(x_i)$ denotes the probability of $x_i$.

\begin{table}[htbp]
\arrayrulecolor{black}
\renewcommand\arraystretch{1.0}
\caption{The min-entropy for different data}
\centering
\small
\begin{tabular}{cc}
\toprule
Variable & Quantum Min-entropy \\
\midrule
$I$ Component & 5.65 bits / 10 bits \\
$Q$ Component& 5.65 bits / 10 bits \\
Reconstructed $\Phi$ & 9.81 bits / 10 bits \\
\bottomrule
\end{tabular}
\end{table}

In our experiment, $n=10$, thus, the probability of each bin is about $P_{\phi_i}=9.77\times{10}^{-4}$. The probability distribution of the phase is illustrated in Fig. 7(c). Submitting the experimental results into Eq. (16), the quantum min-entropy could be estimated for $I$, $Q$, and $\Phi$, which are listed in Table 1. For the original method based on $I$ and $Q$, the quantum min-entropy is about 5.65 bits per sample, but for our phase reconstructed method, the quantum min-entropy with 9.81 bits is achieved. Due to the imperfection of practical devices, the practical quantum min-entropy of our method is slightly lower than the maximal value of 10 bits. 

\begin{figure}[htbp]
\centering
\includegraphics[width=10cm]{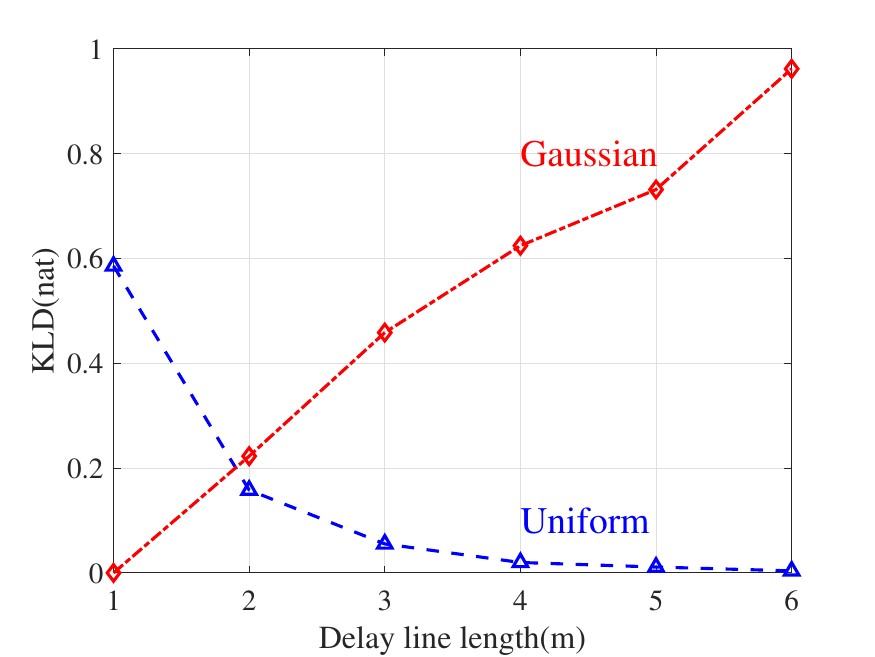}
\caption{Kullback-Leibler Divergence of different delay line lengths. The red rhombus dot-dash line represents the KLD of the reconstructed phase and the standard Gaussian distribution, and the blue triangle dashed line represents the KLD of the reconstructed phase and the standard uniform distribution.}
\end{figure}

Furthermore, to verify the relationship between the distribution of the reconstructed phase and the delay line length of the interferometer, Kullback-Leibler divergence (KLD) is calculated for both Gaussian and uniform distribution under different delay line lengths, which are shown in Fig. 8. For the delay line with length 1m, 2m, 3m, 4m, 5m, and 6m, the phase variances $\langle \Delta \xi(t) \rangle^2$ are about 1.66, 3.32, 5, 6.64, 8.30, and 10, respectively (see Eq. (1)). Figure 8 clearly shows that the KLD rises from 0 to 0.96 when the delay line increases from 1m to 6m. This indicates that the phase gradually deviates from the Gaussian distribution. At the same time, for our reconstructed phase method, the KLD decreases from 0.59 to 0.0039. This implies that, under a 6m delay line, the statistical distribution of the phase can be well approximated by a uniform distribution.

In order to validate the system's robustness against environmental factors, we conducted separate tests to examine the distributions of $I$, $Q$, and $\Phi$ after 30 minutes, 1 hour, and 2 hours of system operation (see Fig. 9). Our findings indicate the system's capability for stable performance over extended durations.

\begin{figure}[htbp]
\centering
\includegraphics[width=13cm]{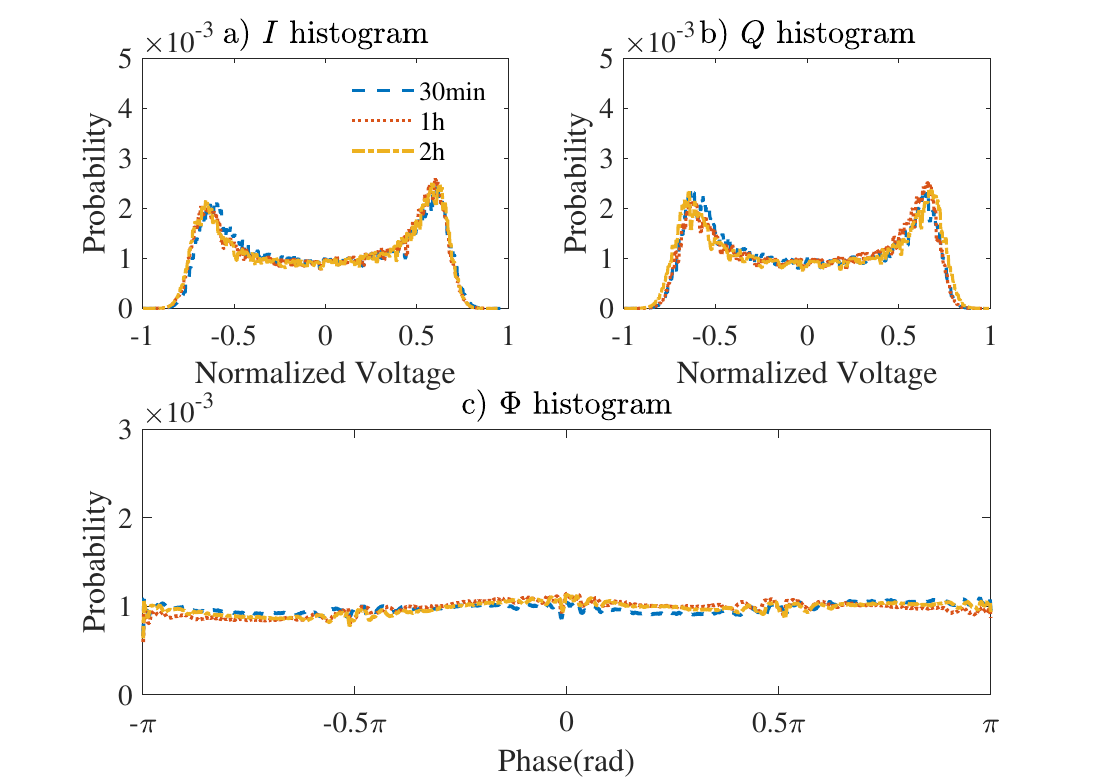}
\caption{Statistic distributions of $I$, $Q$, and $\Phi$ after 30 minutes (the blue dashed line), 1 hour (the red dotted line), and 2 hours (the orange dash-dot line).}
\end{figure}

In general, an elevated data sampling rate results in reduced temporal intervals between adjacent samples, consequently causing stronger correlations. To evaluate the sampling rate in the experiment, the autocorrelation coefficients between adjacent phases are calculated at different sampling rates. The autocorrelation coefficient $R(k)$ of the phase $\Phi$ is defined as
\begin{equation}
    R(k)=\frac{E[(\Phi_{i}-\mu)(\Phi_{i+k}-\mu)]}{\sigma^2}, 
\end{equation}
where $E$ represents the expectation operator, $k$ is the sample delay, $\mu$ and $\sigma^2$ is the mean and the variance of the phase $\Phi$.

\begin{figure}[htbp]
\centering
\includegraphics[width=14cm]{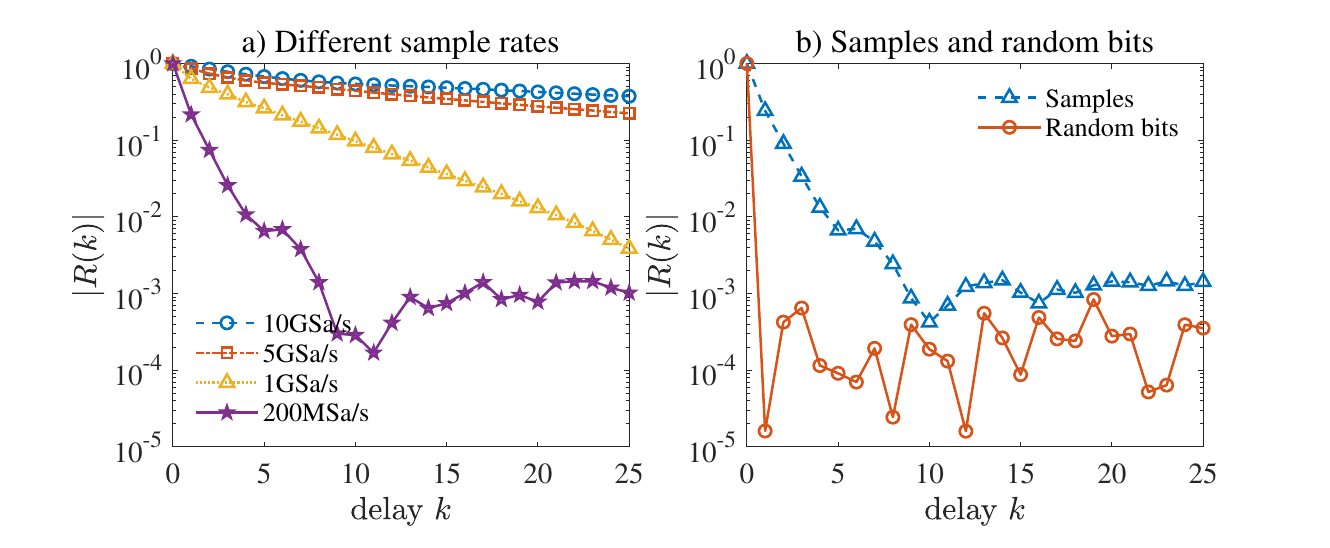}
\caption{Autocorrelation coefficient of the phase. (a) Autocorrelation coefficient of the phase at different sampling rates. The blue circle dashed line represents 10 GSa/s samples; the red square dot-dash line represents 5 GSa/s samples; the orange triangle dotted line represents 1 GSa/s samples; the purple pentagram solid line represents 200 MSa/s samples. (b) The blue triangle dashed line represents the autocorrelation coefficient of the original reconstructed phase samples, and the red circle solid line represents the autocorrelation coefficient of the final random bits after randomness extraction. The autocorrelation coefficient of the final random bits is all less than $1\times{10}^{-3}$.}
\end{figure}

\begin{figure}[htbp]
\centering
\includegraphics[width=13cm]{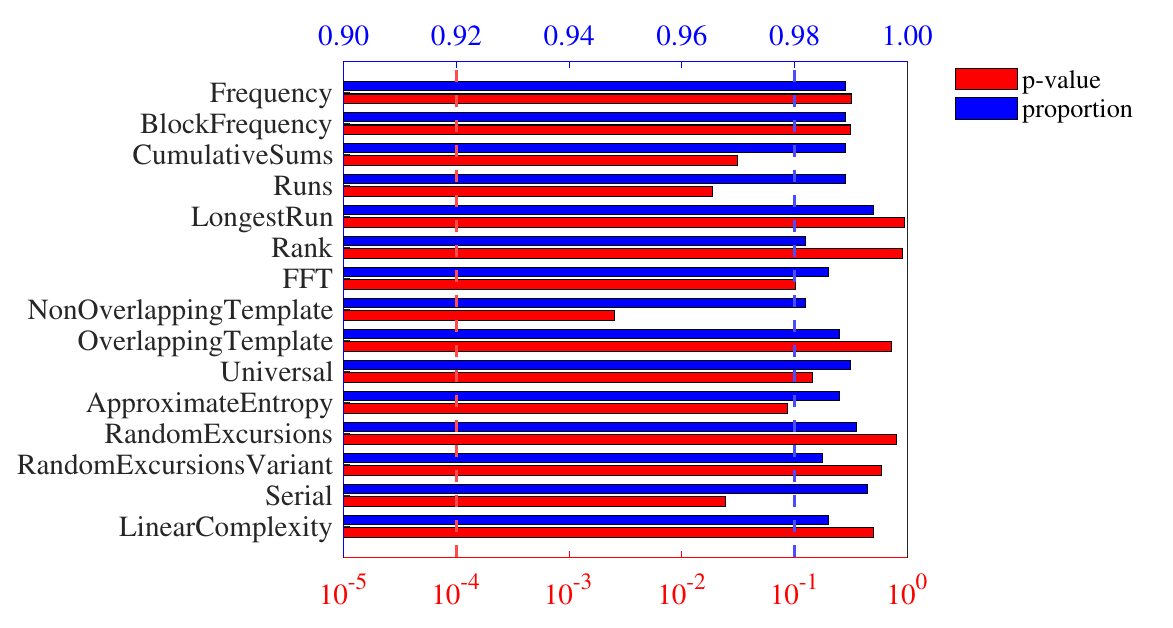}
\caption{Results of the NIST randomness test. The vertical axis in the figure represents 15 test items, the lower horizontal axis represents the p-value, and the upper horizontal axis represents the proportion. The red bars represent the specific p-value of each item; the blue bars represent the specific proportion of each item. It can be seen that the p-value of the 15 tests is greater than $1\times{10}^{-4}$, and the proportion is greater than 0.98.}
\end{figure}

In Fig. 10(a), autocorrelation coefficients $R(k)$ for $1\times{10}^{7}$ phase samples at different sampling rates are shown. The values indicate the level of correlation between the phases as a function of the time lag $k$. When $k=1$, the autocorrelation coefficients are 0.92, 0.85, 0.63, and 0.26 for sampling rates of 10 GSa/s, 5 GSa/s, 1 GSa/s, and 200 MSa/s, respectively. Based on the trade-off between rate and correlation requirements, a sampling rate of 200 MSa/s is suitable for the experiment, which corresponds to a sampling period of $T_s=5$ ns.

Finally, a Toeplitz-hashing randomness extractor is used to compress the original bit sequence and extract pure quantum randomness. The Toeplitz-hashing randomness extractor operates by multiplying the original sequence $n$ by a Toeplitz matrix to generate a random bit sequence $m$. In this experiment, based on the assessment of min-entropy, the size of the Toeplitz matrix was set to $n=4000$ and $m=3920$, which resulted in the generation of approximately 9.8 bits (with a 10-bit ADC) of random bits for each sample. As a result,  a rate of 1.96 Gbps (200 MSa/s×9.8 bits) is achieved.

To evaluate the randomness of the bit sequence from our QRNG, we have examined the autocorrelation coefficients of $10\times{10}^{7}$ bits and subject the sequence to the National Institute of Standards and Technology (NIST) statistical testing suite. The autocorrelation coefficients are below $1\times{10}^{-3}$, and the random bit sequence successfully passes all testing criteria of the NIST statistical testing suite, as depicted in Fig. 10(b) and Fig. 11. This indicates that the generated random bits have good statistical characteristics.

\section{Conclusions}
Random numbers serve as essential components with extensive and vital applications in fields like information security. QRNGs rely on the fundamental principles of quantum physics to generate inherently unpredictable random numbers, significantly enhancing the security of information systems. The QRNGs based on the phase of optical fields have attracted widespread attention due to their simple structure and high generation rates.

A QRNG scheme based on phase reconstruction is proposed in this paper, which provides a way to directly measure the orthogonal quadratures of the optical field from a laser. The proposed scheme offers significant advantages compared to previous QRNGs that rely on phase measurement in unbalanced interferometers, which include achieving a quantum min-entropy approaching 1, improving the efficiency of random number generation, and obtaining a higher random number generation rate under similar conditions. Additionally, the scheme demonstrates robustness against classical phase fluctuations in unbalanced interferometers. The designed QRNG incorporates comprehensive modeling of device imperfections and classical noise, enabling the development of a min-entropy evaluation model that examines the influence of noise on quantum random number performance. The experimental verification successfully achieved a random number generation rate of 1.96 Gbps.

The proposed QRNG can be further improved by increasing the linewidth of the light source to enhance the random number generation rate. However, it’s essential to note that this increased rate comes at the cost of other considerations. In order to prevent the averaging out of random phase within the response time of the detectors, it is crucial to use detectors with higher bandwidth and shorter response time. Therefore, the generation rate of random numbers can be effectively improved by selecting a laser source with higher linewidth \cite{c46} and balanced detectors with higher bandwidth \cite{c38}. 

\begin{backmatter}
\bmsection{Funding}
Shenzhen Science and Technology Program (JCYJ20220818102014029); National Natural Science Foundation of China (62171458); China Electronics Core Research Fund.

\bmsection{Disclosures}
The authors declare that there are no conflicts of interest related to this article.

\bmsection{Data availability}
Data underlying the results presented in this paper are not publicly available at this time but may be obtained from the authors upon reasonable request.

\end{backmatter}


\bibliography{refence}

\begin{thebibliography}{10}
\newcommand{\enquote}[1]{``#1''}

\bibitem{c1}
C.~E. Shannon, \enquote{Communication theory of secrecy systems,}
  {\protect\JournalTitle{The Bell System Technical Journal}} \textbf{28},
  656--715 (1949).

\bibitem{c2}
H.-K. Lo, M.~Curty, and K.~Tamaki, \enquote{Secure quantum key distribution,}
  {\protect\JournalTitle{Nat. Photonics}} \textbf{8}, 595--604 (2014).

\bibitem{c3}
R.~L. Rivest, A.~Shamir, and L.~M. Adleman, \enquote{A method of obtaining
  digital signatures and public-key cryptosystems,}
  {\protect\JournalTitle{Commun. ACM}} \textbf{21}, 120--126 (1978).

\bibitem{c4}
N.~L. Kleinman, J.~C. Spall, and D.~Q. Naiman, \enquote{Simulation-based
  optimization with stochastic approximation using common random numbers,}
  {\protect\JournalTitle{Manage. Sci.}} \textbf{45}, 1570--1578 (1999).

\bibitem{c5}
S.~E. Chick and K.~Inoue, \enquote{New procedures to select the best simulated
  system using common random numbers,} {\protect\JournalTitle{Manage. Sci.}}
  \textbf{47}, 1133--1149 (2001).

\bibitem{c6}
A.~M. Frieze, R.~Kannan, and J.~C. Lagarias, \enquote{Linear congruential
  generators do not produce random sequences,} in \emph{25th Annual Symposium
  on Foundations of Computer Science,}  (1984), pp. 480--484.

\bibitem{c7}
S.~Harase, \enquote{On the f2-linear relations of mersenne twister pseudorandom
  number generators,} {\protect\JournalTitle{Mathematics and Computers in
  Simulation}} \textbf{100}, 103--113 (2014).

\bibitem{c8}
L.~L. Bonilla, M.~Alvaro, and M.~Carretero, \enquote{Chaos-based true random
  number generators,} {\protect\JournalTitle{J. Math. Industry}} \textbf{7}
  (2016).

\bibitem{c9}
Q.~Zhou, X.~Liao, K.~W. Wong, \emph{et~al.}, \enquote{True random number
  generator based on mouse movement and chaotic hash function,}
  {\protect\JournalTitle{Inf. Sci.}} \textbf{179}, 3442--3450 (2009).

\bibitem{c10}
A.~Stefanov, N.~Gisin, O.~Guinnard, \emph{et~al.}, \enquote{Optical quantum
  random number generator,} {\protect\JournalTitle{J. Mod. Opt.}} \textbf{47},
  595--598 (2000).

\bibitem{c11}
T.~Jennewein, U.~Achleitner, G.~Weihs, \emph{et~al.}, \enquote{A fast and
  compact quantum random number generator,} {\protect\JournalTitle{Rev. Sci.
  Instrum.}} \textbf{71}, 1675--1680 (2000).

\bibitem{c12}
M.~A. Wayne \emph{et~al.}, \enquote{Photon arrival time quantum random number
  generation,} {\protect\JournalTitle{J. Mod. Opt.}} \textbf{56}, 516--522
  (2009).

\bibitem{c13}
H.~Xu \emph{et~al.}, \enquote{A spad-based random number generator pixel based
  on the arrival time of photons,} {\protect\JournalTitle{Integration}}
  \textbf{64}, 22--28 (2019).

\bibitem{c14}
Y.-Q. Nie \emph{et~al.}, \enquote{Practical and fast quantum random number
  generation based on photon arrival time relative to external reference,}
  {\protect\JournalTitle{Appl. Phys. Lett.}} \textbf{104}, 051110 (2014).

\bibitem{c15}
M.~Ren \emph{et~al.}, \enquote{Quantum random-number generator based on a
  photon-number-resolving detector,} {\protect\JournalTitle{Phys. Rev. A}}
  \textbf{83}, 023820 (2011).

\bibitem{c16}
Y.~Jian, M.~Ren, E.~Wu, \emph{et~al.}, \enquote{Two-bit quantum random number
  generator based on photon-number-resolving detection,}
  {\protect\JournalTitle{Rev. Sci. Instrum.}} \textbf{82}, 073109 (2011).

\bibitem{c17}
C.~Gabriel \emph{et~al.}, \enquote{A generator for unique quantum random
  numbers based on vacuum states,} {\protect\JournalTitle{Nat. Photonics}}
  \textbf{4}, 711--715 (2010).

\bibitem{c18}
J.~G. Raity, P.~C.~M. Owens, and P.~R. Tapster, \enquote{Quantum random-number
  generation and key sharing,} {\protect\JournalTitle{J. Mod. Opt.}}
  \textbf{41}, 2435--2444 (1994).

\bibitem{c52}
Z.~Zheng, Y.~Zhang, W.~Huang, \emph{et~al.}, \enquote{6 gbps real-time optical
  quantum random number generator based on vacuum fluctuation,}
  {\protect\JournalTitle{Review of Scientific Instruments}} \textbf{90} (2019).

\bibitem{c19}
B.~Qi \emph{et~al.}, \enquote{High-speed quantum random number generation by
  measuring phase noise of a single-mode laser,} {\protect\JournalTitle{Opt.
  Lett.}} \textbf{35}, 312--314 (2010).

\bibitem{c20}
C.~Abell{\'a}n \emph{et~al.}, \enquote{Ultra-fast quantum randomness generation
  by accelerated phase diffusion in a pulsed laser diode,}
  {\protect\JournalTitle{Opt. Express}} \textbf{22}, 1645--1654 (2014).

\bibitem{c21}
Y.-Q. Nie \emph{et~al.}, \enquote{The generation of 68 gbps quantum random
  number by measuring laser phase fluctuations,} {\protect\JournalTitle{Rev.
  Sci. Instrum.}} \textbf{86}, 063105 (2015).

\bibitem{c22}
M.~Huang, Z.~Chen, Y.~Zhang, and H.~Guo, \enquote{A phase fluctuation based
  practical quantum random number generator scheme with delay-free structure,}
  {\protect\JournalTitle{Appl. Sci.}} \textbf{10}, 2431 (2020).

\bibitem{c23}
F.~Xu \emph{et~al.}, \enquote{Ultrafast quantum random number generation based
  on quantum phase fluctuations,} {\protect\JournalTitle{Opt. Express}}
  \textbf{20}, 12366--12377 (2012).

\bibitem{c49}
M.~Jofre \emph{et~al.}, \enquote{True random numbers from amplified quantum
  vacuum,} {\protect\JournalTitle{Opt. Express}} \textbf{19}, 20665 (2011).

\bibitem{c51}
J.~Yang, M.~Wu, Y.~Zhang, \emph{et~al.}, \enquote{An ultra-fast quantum random
  number generation scheme based on laser phase noise,}
  {\protect\JournalTitle{arXiv preprint arXiv:2311.17380}}  (2023).

\bibitem{c24}
J.~Yang \emph{et~al.}, \enquote{Randomness quantification for quantum random
  number generation based on detection of amplified spontaneous emission
  noise,} {\protect\JournalTitle{Quantum Sci. Technol.}} \textbf{6}, 015002
  (2021).

\bibitem{c25}
C.~R.~S. Williams \emph{et~al.}, \enquote{Fast physical random number generator
  using amplified spontaneous emission,} {\protect\JournalTitle{Opt. Express}}
  \textbf{18}, 23584 (2010).

\bibitem{c30}
T.~Bertapelle \emph{et~al.}, \enquote{High-speed source-device-independent
  quantum random number generator on a chip,} {\protect\JournalTitle{arXiv}}
  (2023).

\bibitem{c47}
T.~Roger \emph{et~al.}, \enquote{Interferometric quantum random number
  generation on chip,} {\protect\JournalTitle{2019 Conference on Lasers and
  Electro-Optics (CLEO)}} pp. 1--2 (2019).

\bibitem{c48}
B.~Bai \emph{et~al.}, \enquote{18.8 gbps real-time quantum random number
  generator with a photonic integrated chip,} {\protect\JournalTitle{Appl.
  Phys. Lett.}} \textbf{118}, 264001 (2021).

\bibitem{c26}
Y.~Liu \emph{et~al.}, \enquote{Device-independent quantum random-number
  generation,} {\protect\JournalTitle{Nature}} \textbf{562}, 548--551 (2018).

\bibitem{c27}
D.~G. Marangon, G.~Vallone, and P.~Villoresi,
  \enquote{Source-device-independent ultrafast quantum random number
  generation,} {\protect\JournalTitle{Phys. Rev. Lett.}} \textbf{118}, 060503
  (2017).

\bibitem{c28}
M.~H. Li \emph{et~al.}, \enquote{Experimental realization of device-independent
  quantum randomness expansion,} {\protect\JournalTitle{Phys. Rev. Lett.}}
  \textbf{126}, 050503 (2021).

\bibitem{c29}
J.~Cheng \emph{et~al.}, \enquote{Mutually testing source-device-independent
  quantum random number generator,} {\protect\JournalTitle{Photonics Res.}}
  \textbf{10}, 646--652 (2022).

\bibitem{c31}
T.~Song, X.~Tan, and J.~Weng, \enquote{Statistical fluctuation analysis of
  measurement-device-independent quantum random-number generation,}
  {\protect\JournalTitle{Phys. Rev. A}} \textbf{99}, 022333 (2019).

\bibitem{c32}
Y.-Q. Nie \emph{et~al.}, \enquote{Experimental measurement-device-independent
  quantum random-number generation,} {\protect\JournalTitle{Phys. Rev. A}}
  \textbf{94}, 060301 (2016).

\bibitem{c33}
M.~Herrero-Collantes and J.~C. Garc{\'i}a-Escart{\'i}n, \enquote{Quantum random
  number generators,} {\protect\JournalTitle{Rev. Mod. Phys.}} \textbf{89},
  015004 (2017).

\bibitem{c34}
H.~Guo \emph{et~al.}, \enquote{Truly random number generation based on
  measurement of phase noise of a laser,} {\protect\JournalTitle{Phys. Rev. E}}
  \textbf{81}, 051137 (2010).

\bibitem{c35}
A.~Yariv and P.~Yeh, \emph{Photonics: optical electronics in modern
  communications} (Oxford university press, 2007).

\bibitem{c36}
D.~Collett and T.~Lewis, \enquote{Discriminating between the von mises and
  wrapped normal distributions,} {\protect\JournalTitle{Aust. J. Stat.}}
  \textbf{23}, 73--79 (1981).

\bibitem{c37}
G.~Kurz, I.~Gilitschenski, and U.~D. Hanebeck, \enquote{Efficient evaluation of
  the probability density function of a wrapped normal distribution,} in
  \emph{2014 Sensor Data Fusion: Trends, Solutions, Applications (SDF),}
  (2014), pp. 1--5.

\bibitem{c38}
X.~G. Zhang \emph{et~al.}, \enquote{68 gbps quantum random number generation by
  measuring laser phase fluctuations,} {\protect\JournalTitle{Rev. Sci.
  Instrum.}} \textbf{86}, 063105 (2015).

\bibitem{c39}
S.-H. Sun and F.~Xu, \enquote{Experimental study of a quantum random-number
  generator based on two independent lasers,} {\protect\JournalTitle{Phys. Rev.
  A}} \textbf{96}, 062314 (2017).

\bibitem{c40}
C.~Abellan \emph{et~al.}, \enquote{Quantum entropy source on an inp photonic
  integrated circuit for random number generation,}
  {\protect\JournalTitle{Optica}} \textbf{3}, 989--994 (2016).

\bibitem{c41}
T.~Roger \emph{et~al.}, \enquote{Real-time interferometric quantum random
  number generation on chip,} {\protect\JournalTitle{J. Opt. Soc. Am. B}}
  \textbf{36}, B137 (2019).

\bibitem{c42}
B.~Septriani \emph{et~al.}, \enquote{Parametric study of the phase diffusion
  process in a gain-switched semiconductor laser for randomness assessment in
  quantum random number generator,} {\protect\JournalTitle{AIP Advances}}
  \textbf{10}, 055022 (2020).

\bibitem{c43}
J.-R. Álvarez \emph{et~al.}, \enquote{Random number generation by coherent
  detection of quantum phase noise,} {\protect\JournalTitle{Opt. Express}}
  \textbf{28}, 5538--5547 (2020).

\bibitem{c44}
Y.~Painchaud \emph{et~al.}, \enquote{Performance of balanced detection in a
  coherent receiver,} {\protect\JournalTitle{Opt. Express}} \textbf{17},
  3659--3672 (2009).

\bibitem{c45}
R.~Shakhovoy \emph{et~al.}, \enquote{Phase randomness in a semiconductor laser:
  Issue of quantum random-number generation,} {\protect\JournalTitle{Phys. Rev.
  A}} \textbf{107}, 012616 (2023).

\bibitem{c50}
X.~Ma, F.~Xu, H.~Xu, \emph{et~al.}, \enquote{Postprocessing for quantum
  random-number generators: Entropy evaluation and randomness extraction,}
  {\protect\JournalTitle{Physical Review A}} \textbf{87}, 062327 (2013).

\bibitem{c46}
J.~Yang \emph{et~al.}, \enquote{Randomness quantification for quantum random
  number generation based on detection of amplified spontaneous emission
  noise,} {\protect\JournalTitle{Quantum Sci. Technol.}} \textbf{6}, 015002
  (2021).

\end{thebibliography}

\end{document}